%% file: main.tex
\documentclass[acmsmall,screen]{acmart}
\usepackage[utf8]{inputenc}
\usepackage{booktabs}   
\usepackage{multirow}   
\usepackage{graphicx}   
\usepackage{arydshln}   
\usepackage{subcaption} 
\usepackage{multirow}
\usepackage{array}      
\usepackage{makecell}
\usepackage{threeparttable}
\AtBeginDocument{%
  }
\usepackage{tcolorbox}
\usepackage{colortbl}
\usepackage{fancyvrb}
\usepackage{xcolor}
\usepackage{pifont}

\definecolor{darkgreen}{rgb}{0,0.5,0} 
\definecolor{purple}{RGB}{90,60,150} 
\definecolor{todocolor}{rgb}{0.9,0.1,0.1} 
\definecolor{fixcolor}{rgb}{0.1,0.7,0.3} 
\definecolor{hycolor}{rgb}{0.7,0.7,0.3} 
\definecolor{wycolor}{rgb}{0.9,0.1,0.1} 

\definecolor{lightred}{RGB}{255,225,220}
\definecolor{lightblue}{RGB}{52,119,203}
\definecolor{softyellow}{RGB}{200,140,20} 

\definecolor{summarylightblue}{RGB}{243,248,252}
\newenvironment{summary}{
\begin{tcolorbox}[width=\linewidth, colback=summarylightblue, top=1pt, bottom=1pt, left=2pt, right=2pt]

}
{
\end{tcolorbox}
}

\definecolor{tablelightblue}{RGB}{255,225,220} 

\newcount\DraftStatus  
\newcommand{\nbc}[3]{\ifnum\DraftStatus=1
	{\colorbox{#3}{\bfseries\sffamily\scriptsize\textcolor{white}{#1}}}
	{\textcolor{#3}{\sf\small$\blacktriangleright$\emph{#2}$\blacktriangleleft$}}
\fi}

\newcommand{\draftnote}[2]{\ifnum\DraftStatus=1
	\marginpar{
		\tiny\raggedright
		\hbadness=10000
		\def\baselinestretch{0.8}
		\textcolor{#1}{\textsf{\hspace{0pt}#2}}}
\fi}

\usepackage{babel}
\usepackage{hyperref}
\usepackage{caption}
\usepackage{xspace}

\definecolor{lightcyan}{RGB}{10,110,150}
\hypersetup{
  colorlinks=true,
  linkcolor=lightcyan,
  citecolor=lightcyan,
  filecolor=lightcyan,
  urlcolor=lightcyan
}
\usepackage[capitalize,noabbrev]{cleveref}

\newtcbox{\capsule}[1]{%
  on line,
  colback=#1!15,
  colframe=#1!60!black,
  arc=2pt,
  boxrule=0.3pt,
  left=1.pt,right=1.5pt,
  top=0.1pt,bottom=0.1pt,
}

\setcopyright{acmlicensed}
\copyrightyear{2026}
\acmYear{2026}
\begin{document}
\title{HerAgent: Rethinking the Automated Environment Deployment via Hierarchical Test Pyramid}


\author{Xiang Li}
\affiliation{%
  \institution{University College London}
  \city{London}
  \country{United Kingdom}}
\email{x.li.25@ucl.ac.uk}

\author{Siyu Lu}
\affiliation{%
  \institution{Uppsala University}
  \city{Uppsala}
  \country{Sweden}
}
\email{siyu.lu.6562@student.uu.se}

\author{Federica Sarro}
\affiliation{%
  \institution{University College London}
  \city{London}
  \country{United Kingdom}}
\email{f.sarro@ucl.ac.uk}

\author{Claire Le Goues}
\affiliation{%
  \institution{Carnegie Mellon University}
  \city{Pittsburgh}
  \country{United States}
}
\email{clegoues@cs.cmu.edu}

\author{He Ye}
\affiliation{%
  \institution{University College London}
  \city{London}
  \country{United Kingdom}}
\email{he.ye@ucl.ac.uk}

\renewcommand{\shortauthors}{Xiang Li et al.}
\newcommand{\OurApproach}{\textit{HerAgent}\xspace}
\newcommand{\bashfile}{\textit{Bash File}\xspace}
\newcommand{\testpyramid}{\textit{Test Pyramid}\xspace}

\newcommand{\maturityI}{\textcolor{lightblue}{Installability}\xspace}
\newcommand{\maturityII}{\textcolor{hycolor}{Testability}\xspace}
\newcommand{\maturityIII}{\textcolor{purple}{Runnability}\xspace}

\begin{abstract}
Automated software environment setup is a prerequisite for testing, debugging, and reproducing failures, yet remains challenging in practice due to complex dependencies, heterogeneous build systems, and incomplete documentation. Recent work leverages large language models to automate this process, but typically evaluates success using weak signals such as dependency installation or partial test execution, which do not ensure that a project can actually run.

In this paper, we argue that environment setup success should be evaluated through executable evidence rather than a single binary signal. We introduce the Environment Maturity Hierarchy, which defines three success levels based on progressively stronger execution requirements, culminating in successful execution of a project’s main entry point. 

Guided by this hierarchy, we propose \OurApproach, an automated environment setup approach that incrementally constructs executable environments through execution-based validation and repair. We evaluate \OurApproach on four public benchmarks, where it outperforms all related work, achieving up to 79.6\% improvement due to its holistic understanding of project structure and dependencies. On complex C/C++ projects, \OurApproach surpasses prior approaches by 66.7\%. In addition, \OurApproach uniquely resolves 11–30 environment instances across the benchmarks that no prior method can configure.

\end{abstract}




\maketitle

\section{Introduction}

\input{sections/1.Introduction}

\section{Problem Statement}
\label{sec:problem_statement}

\input{sections/2.ProblemStatement}

\section{Approach}

\input{sections/3.Approach}

\section{Experimental Setup}

\input{sections/4.ExperimentalSetup}

\section{Experimental Results}

\input{sections/5.ExperimentResults}

\section{Related Work}
\input{sections/6.RelatedWorks}

\section{Conclusion}
In this paper, we addressed the ambiguity of success criteria and the limitations of partial verification signals in automated software environment setup. We introduced the Environment Maturity Hierarchy, a formal framework that redefines setup success by distinguishing between Installable, Testable, and Runnable states, thereby establishing a rigorous standard for "User-Ready" environments. To operationalize this, we proposed \OurApproach, a multi-agent system that leverages a Script-centric Repair mechanism. By persisting configuration states in a global script rather than relying on ephemeral command executions, \OurApproach robustly handles the complexity of heterogeneous repositories. Our extensive evaluation across four benchmarks demonstrates that \OurApproach significantly outperforms state-of-the-art baselines. Crucially, our empirical results reveal a distinct gap between the Testable and Runnable states, confirming that passing unit tests is an insufficient proxy for end-to-end system usability. In summary, \OurApproach not only advances the state of the art in automated environment configuration but also provides a foundational capability for future autonomous software engineering agents to validate and execute code in realistic, user-ready environments.

\section{Data Availability} Our experimental data and source code are available at ~\url{https://github.com/EuniAI/EnvAgent}.

\bibliographystyle{ACM-Reference-Format}
\bibliography{references}


\end{document}

%% file: sections/1.Introduction.tex
Automated software environment deployment \cite{hu2025repo2run,bouzenia2025you,milliken2025beyond,kovrigin2025piper,arora2025setupbench,guo2025swe} is a necessary step to smoothly run software projects. It includes installing dependencies, building the project, and running test cases to ensure that a software works in a given environment. Correct environment deployment is required for tasks such as debugging \cite{rewardrepair}, testing \cite{xie2025repost}, and reproducing failures \cite{majgaonkar2025understandingcodeagentbehaviour,wang2025swe}.

The drive to automate project environment deployment stems from two primary motivations. First, it mitigates the friction of manual configuration caused by complex stacks and outdated documentation. Empirical studies underscore this challenge, revealing that 38–60\% of Java builds fail in simulated or variant environments due to reproducibility issues~\cite{Sul_r_2016, sulír2024localsoftwarebuildabilityjava}. Furthermore, automated solutions like CI/CD are essential to resolve the "reproducibility crisis," ensuring consistent execution contexts over vague manual procedures~\cite{beaulieu2017reproducibility}.
Second, it is a prerequisite for coding agents \cite{repairAgent-icse25,chen2025prometheusunifiedknowledgegraphs,traeresearchteam2025traeagent,autocoderrover} to first build executable environments to provide feedback via compilation and testing. Without a functional environment, agents cannot validate changes or observe failures, making automated deployment a cornerstone of agent-based software engineering.

Recent advances in large language models (LLMs) have significantly improved automated environment deployment. In these approaches, LLMs are mainly responsible for reasoning about required dependencies and test execution commands \cite{milliken2025beyond, bouzenia2025you}. Prior work explores building Docker containers \cite{hu2025repo2run, hu2025compileagent} or generating shell scripts \cite{kovrigin2025piper} to automatically deploy projects and evaluate results through static analysis or test-suite execution.

The prior works are promising, but incomplete.  Existing approaches to automated software environment deployment exhibit three key limitations:

\textbf{Problem 1: Prior work defines environment deployment success with different and incomplete criteria.}
Prior work equates environment deployment success with successful builds, static analysis (e.g, PIPER \cite{kovrigin2025piper} and EnvBench \cite{eliseeva2025envbench}, or the ability to invoke test frameworks e.g., Repo2Run \cite{hu2025repo2run} and Installmatic \cite{milliken2025beyond}. Such criteria do not validate the program’s main execution entry point. In practice, human developers \cite{4222596} and coding agents \cite{yang2024sweagent,wang2025openhands,autocoderrover} expect to know how to run the program.

\textbf{Problem 2: Prior work lacks a holistic understanding of projects that leads to limited effectiveness and poor scalability.}
Prior work typically constructs environment by reacting to individual compilation or test errors, rather than reasoning about the repository as a whole. All prior work \cite{milliken2025beyond,bouzenia2025you,hu2025repo2run,kovrigin2025piper} start to download missing dependencies are added only after a compilation failure is observed, without considering the project’s dependency graph or execution workflow. As a result, environment deployment is driven by local failure signals, leading to repeated errors, brittle fixes, and limited scalability to complex or heterogeneous repositories.

\textbf{Problem 3: Prior work relies on strong assumptions about specific project structure and small scale evaluations, reduce external validity.}
Prior work is typically designed for projects with fixed repository structures or predefined environment construction patterns, such as Python projects organized around \texttt{pytest} \cite{hu2025repo2run,milliken2025beyond} or workflows based on \texttt{pyright} \cite{kovrigin2025piper,eliseeva2025envbench}. As a result, evaluations are often conducted on small size. For example, ExecutionAgent \cite{bouzenia2025you} was evaluated on 50 instances and Installmatic \cite{milliken2025beyond} was evaluated on 40 instances. 
Given the wide diversity of real-world codebases in language, structure, and execution workflows, the external validity of these approaches remains unclear.

\begin{figure}[t]
  \centering
  \includegraphics[width=\columnwidth]{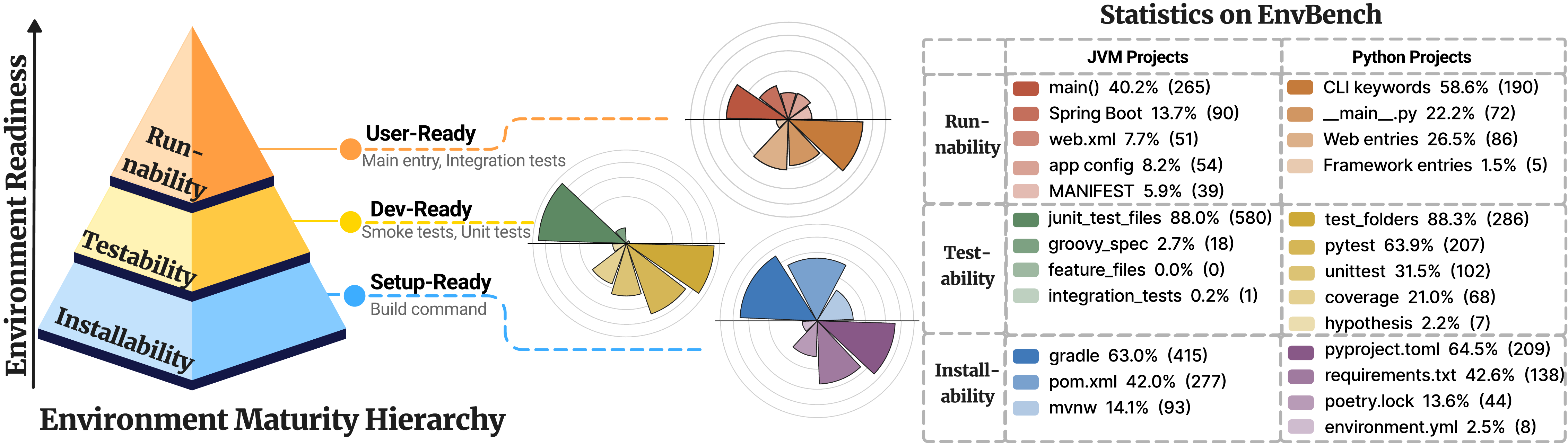}
  \caption{The Environment Maturity Hierarchy and Ecosystem Reality. Left: The three-stage maturity model. Right: Command distribution across 659 JVM and 324 Python repositories in EnvBench~\cite{eliseeva2025envbench}. The rose charts show the rich diversity of dependencies and test suites at the Installable, Testable, and Runnable levels.}
  \label{fig:hierarchy}
\end{figure}

\textbf{Our Solution - \OurApproach:} 
To address the above problems, we propose \OurApproach, a novel automated framework for software environment deployment. \underline{To address Problem 1}, \OurApproach introduces an \emph{Environment Maturity Hierarchy} that defines three success signals for environment deployment: Installability, Testability, and Runnability. This hierarchy explicitly identifies the project’s main execution entry point to ensure the project can be run. This is more challenging than simply executing commands such as \texttt{pytest} or \texttt{mvn test}.
\underline{To address Problem 2}, \OurApproach first analyzes the project structure using a knowledge graph to obtain a holistic understanding of project components and dependencies. This design is fundamentally different from prior work and enables \OurApproach to handle complex projects more effectively.
\underline{To address Problem 3}, \OurApproach is designed to generalize across diverse project structures and is not tied to specific repository conventions, such as pytest-centric layouts.

\OurApproach is extensively evaluated on four automated environment deployment benchmarks: EnvBench~\cite{eliseeva2025envbench}, Repo2Run-Bench~\cite{hu2025repo2run}, ExecutionAgent-Bench~\cite{bouzenia2025you}, and Installamatic-Bench~\cite{milliken2025beyond}. The evaluation covers 14 different programming languages and includes comparisons with four related approaches (PIPER~\cite{kovrigin2025piper}, ExecutionAgent~\cite{bouzenia2025you}, Repo2Run~\cite{hu2025repo2run}, and Installamatic~\cite{milliken2025beyond}), as well as frontier closed-source models (GPT series) and open-source models (Qwen series). 

Our experimental results show that \OurApproach outperforms all related work across all four benchmarks, achieving up to a 79.6\% improvement. This performance gain is attributed to the holistic understanding of project structure and dependencies enabled by \OurApproach. In particular, for complex C and C++ projects, \OurApproach outperforms 66.7\% of prior approaches.

In summary, our contributions are as follows:
\begin{itemize}

\item \textbf{Conceptual Novelty:} We propose the \emph{Hierarchy Test Pyramid}, which systematically defines three evaluation metrics to characterize the success levels of automated project environment deployment. To the best of our knowledge, this is the first work to provide a unified success definition in this area.

\item \textbf{Approach Novelty:} We propose \OurApproach, which holistically understands project structure and dependencies directly from codebases, instead of relying on reactive, error-driven environment construction. This holistic design enables \OurApproach to effectively handle complex projects, particularly in C and C++, outperforming prior work by up to 66.7\%.

\item \textbf{State-of-the-art Performance:} We conduct a comprehensive evaluation of \OurApproach on four automated environment deployment benchmarks and demonstrate state-of-the-art performance across all benchmarks, compared with representative prior approaches as well as frontier closed-source and open-source models.

\item \textbf{Artifact Availability:} We release all code, execution trajectories, and experimental results to facilitate reproducibility and support future research~\url{https://github.com/EuniAI/EnvAgent}.

\end{itemize}

%% file: sections/2.ProblemStatement.tex
\subsection{Environment Maturity Hierarchy}
\label{sec:Hierarchy}

We first manually inspect 659 JVM and 324 Python projects from EnvBench~\cite{eliseeva2025envbench} to examine how build, test, and execution commands are used in real repositories (Fig.~\ref{fig:hierarchy}, right). This manual analysis reveals a clear stratification of commands in practice: dependency installation, test execution, and running the application exercise the system to very different extents and provide different levels of confidence about whether a project can actually run.

Motivated by this observation, we introduce the \emph{Environment Maturity Hierarchy} (Fig.~\ref{fig:hierarchy}, left), which organizes executable commands into three levels based on the evidence they provide about environment readiness. Rather than treating all tests equally, the hierarchy forms a \testpyramid, where higher levels correspond to executions closer to real usage and thus offer stronger assurance. We now introduce these three success metrics for automated environment deployment.

\maturityI: This metric indicates whether declared dependencies can be successfully installed under a given platform and toolchain. An environment is considered \emph{setup-ready} if it can execute build or installation commands, such as \texttt{mvn install} or \texttt{pip install -r requirements.txt}. Even at this basic level, repositories exhibit substantial diversity in build systems and dependency specifications across JVM and Python projects. While these commands confirm that dependencies are installable, this level provides no evidence that the program can run at runtime.

\maturityII: An environment satisfies this metric if it can execute test-oriented commands, such as smoke tests or unit tests (e.g., \texttt{pytest}, \texttt{mvn test}, or simple \texttt{--version} checks). These commands invoke the runtime and provide limited evidence that the environment is functional. In practice, repositories expose diverse testing commands and structures, and tests often rely on mocks or avoid full execution paths.  This level provides limited evidence that the program can run correctly.  Most prior work focuses on this level and treats it as success.

\maturityIII: 
This level is reached when the program’s main entry point or integrated workflows can be executed successfully, such as running \texttt{python main.py}, launching a command-line interface, or executing integration tests that interact with external services. While these execution commands are diverse, they share a common property: they exercise the system end to end under realistic conditions. Successful execution at this level indicates that dependencies, configuration, and component interactions work together correctly. We therefore treat main-entry execution and integration tests as equivalent evidence of reaching this state. This level represents a fully configured environment.

\subsection{Formal Definition of Success}

The \emph{Environment Maturity Hierarchy} provides a conceptual view of progressive environment readiness. To clearly describe the process of switching between different states, we now formalize this notion by defining state-aware success criteria

Let $\mathcal{S} = \{\mathsf{Installability}, \mathsf{Testability}, \mathsf{Runnability}\}$ denote the set of environment maturity states. These states are partially ordered by increasing execution guarantees:
\[
\mathsf{Installability} \subsetneq \mathsf{Testability} \subsetneq \mathsf{Runnability}
\]
This relation denotes a hierarchical dependency: achieving a higher state inherently necessitates the capabilities of the lower states (necessity), whereas possessing the capabilities of a lower state is insufficient to guarantee the higher state (non-sufficiency). For instance, an environment must be installable to be testable, but successful installation does not imply test capability.

For each state $s \in \mathcal{S}$, we associate a \testpyramid  $C_s$, a set of validation commands. 
Each command $c \in C_s$ functions as an executability oracle: its successful execution provides concrete evidence that the environment satisfies the criteria of state $s$. Concretely:  
\begin{itemize}
    \item $C_{\mathsf{Installability}}$ includes build and installation commands in \maturityI (e.g., \texttt{pip install -r requirements.txt}, \texttt{npm install}),
    \item $C_{\mathsf{Testability}}$ includes unit tests, smoke tests, and simple runtime probes in \maturityII (e.g., \texttt{pytest}, \texttt{--version} commands),
    \item $C_{\mathsf{Runnability}}$ includes main entry-point executions and integration tests that exercise full application workflows in \maturityIII (e.g., \texttt{python main.py}, CLI startup commands, or end-to-end integration tests).
\end{itemize}

We formalize command execution using a binary oracle function $\mathit{exec}(\cdot)$:
\[
\mathit{exec}(c) = 
\begin{cases}
1, & \text{if command $c$ terminates successfully (return code 0)},\\
0, & \text{otherwise}.
\end{cases}
\]
The aggregated outcome of all commands in a state $s$ is denoted as $\mathit{exec}(C_s)$, defined such that
\[
\mathit{exec}(C_s) = 1 \quad \Longleftrightarrow \quad \exists\, c \in C_s \text{ s.t. } \mathit{exec}(c) = 1.
\]
In this definition, a single successful validation is sufficient evidence that the environment has achieved the corresponding maturity level.

With these definitions in place, we can formalize environment state transitions. Let $\delta(s)$ denote the state-aware transition function, which determines the next maturity state of the environment based on the outcome of validation commands:
\[
\delta(s) =
\begin{cases}
s^+, & \mathit{exec}(C_s) = 1, \\
s^-, & \mathit{exec}(C_s) = 0,
\end{cases}
\quad \text{with } s^- \subsetneq s \subsetneq s^+.
\]
Here, $s^+$ represents the next higher maturity state, indicating that the environment can safely progress, while $s^-$ represents the next lower state, capturing regression due to failed commands. $\delta(s)$ is not a fixed rule-based automaton, but a prompt-driven agent policy that decides whether to advance, remain, or rollback the current maturity state based on execution outputs and the current \testpyramid. $s$ is allowed to progress to higher maturity when validation succeeds, while rollback to a lower state can be applied if significant errors occur.

Finally, we define $s^\star$ as the maximum supported maturity state of an environment as the highest state in $\mathcal{S}$ for which at least one associated validation command succeeds:
\[
s^\star = \max \{ s \in \mathcal{S} \mid \mathit{exec}(C_s) = 1 \}.
\]
In practice, $s^\star$ serves as a benchmark for automated environment validation and repair, guiding both developers and autonomous agents in targeting specific maturity objectives. By leveraging the $\mathit{exec}(C_s)$ oracle and the associated transition function $\delta(s)$, we can reason systematically about both incremental improvement and failure handling in automated environment deployment.

%% file: sections/3.Approach.tex
\begin{figure}[t]
  \centering
  \includegraphics[width=\columnwidth]{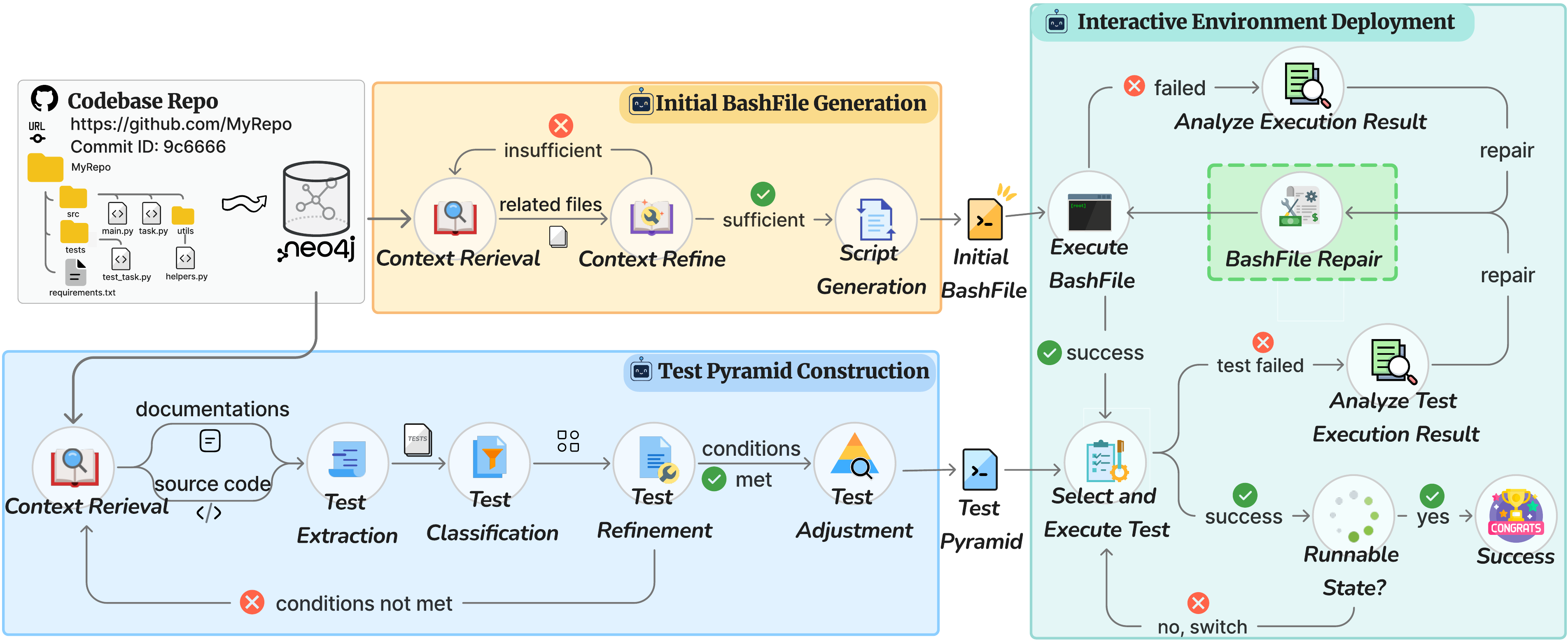}
  \caption{Overview of \OurApproach. The pipeline comprises: (1) \bashfile Generation (yellow) to construct a initial script; (2) Test Pyramid Construction (blue) to retrieve and categorize test commands into the hierarchy; and (3) Interactive Environment Deployment (green), where a dual-loop repair mechanism iteratively validates and advances environment maturity.}
  \label{fig:overview}
\end{figure}

We propose \OurApproach, an automated environment setup agent. \OurApproach is designed around the \emph{Environment Maturity Hierarchy} and implemented as a multi-agent system. As shown in Figure~\ref{fig:overview}, the workflow is divided into three stages: 1) Initial BashFile Generation (yellow), 2) Test Pyramid Construction (blue), and 3) Interactive Environment Deployment (green).

\OurApproach takes as input a repository (i.e., a GitHub URL or a local codebase). It automatically generates a runnable \bashfile, identifies the relevant test commands required by the \testpyramid, and iteratively constructs an executable environment that satisfies the success criteria defined by the hierarchy.

As output, \OurApproach produces a validated execution specification, including a runnable \bashfile and a corresponding Dockerfile that materializes the environment.

\subsection{Initial BashFile Generation}

To bootstrap automated environment deployment, we first construct an initial executable environment configuration in \bashfile (see yellow area in Fig.~\ref{fig:overview}). The goal of this phase is not to produce a fully correct or optimal environment, but to establish a concrete and executable starting point that reflects the repository’s declared dependencies, tooling conventions, and build assumptions. We adopt an iterative three-step procedure to generate the initial version of \bashfile: \emph{Retrieval}, \emph{Refine}, and \emph{Generation}.

\textbf{Env-Context Retrieval and Refinement}. Follow Prometheus~\cite{chen2025prometheusunifiedknowledgegraphs}, \OurApproach captures the entire codebase using a knowledge graph based on Tree-sitter~\footnote{\url{https://tree-sitter.github.io/tree-sitter/}} and Neo4J~\footnote{\url{https://neo4j.com}}, enabling the agent to navigate and search for environment configuration-related files, and then summarize, analyze, and categorize them. Many tools that search for files in different ways will be used, including file lookup, documentation traversal, file preview tools. During iterative retrieval, \OurApproach records an \texttt{involved\_files} list and injects it into subsequent prompts, to avoid repeatedly querying the same files across iterations.

After obtaining the relevant files obtained through the search, the file-tree information are fed into the Agent, to determine whether the found files are sufficient to generate a \bashfile. If the files are analyzed insufficient, \OurApproach generates search suggestions and return to the previous Retrieval step; otherwise if the files are analyzed to be sufficient or the searching reaches the maximum query round, it will stop the search and enter the next Generation step.

\textbf{BashFile Generation}. After obtaining a sufficient number of environment configuration-related files, \OurApproach can call tools (e.g., write and read tool) to analyze the contents of  relevant files.

\begin{figure}[t]
  \centering
  \includegraphics[width=\columnwidth]{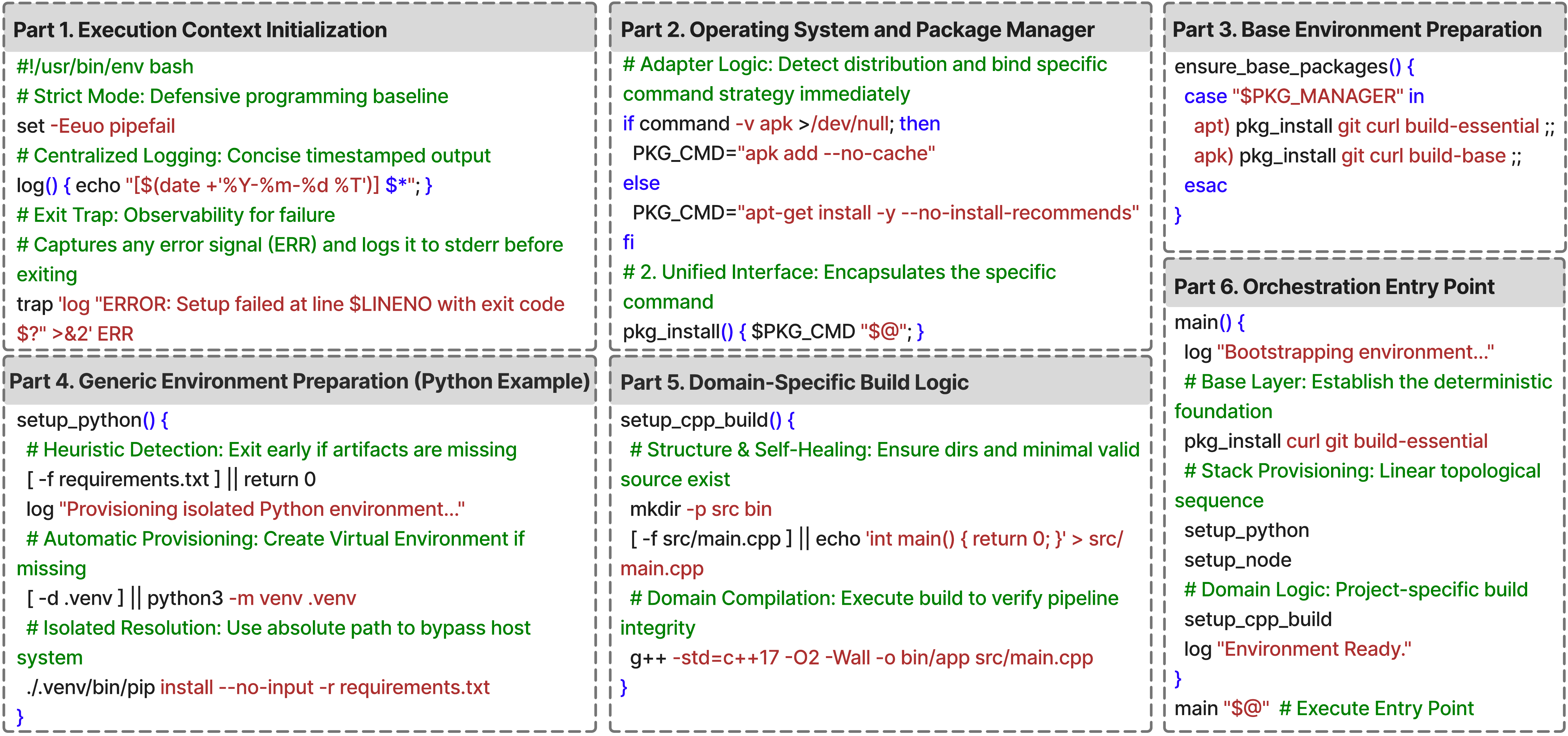}
  \caption{An example of the \bashfile template used in \OurApproach consists of six steps. This \bashfile can be executed directly in an isolated sandbox container.}
  \label{fig:bashfile}
\end{figure}

Figure~\ref{fig:bashfile} shows an example of a structured \bashfile template consisting of six steps.
\OurApproach generates a \bashfile that can be executed directly in an isolated sandbox container. To ensure robustness and reproducibility, the template is organized into logical layers:

\begin{enumerate}
    \item Execution Context Initialization. Enforces strict execution semantics (e.g., \texttt{set -e}) and enables centralized logging to ensure system observability and robust error handling.
    \item OS and Package Manager Abstraction. Unifies package manager commands (e.g., \texttt{apt}, \texttt{apk}) via an adapter interface to guarantee cross-platform portability.
    \item Base Environment Preparation. Installs essential system utilities (e.g., \texttt{git}, compilers) to provide a consistent and reproducible foundation across environments.
    \item Generic Environment Preparation. Provisions isolated runtimes based on detected language artifacts (e.g., \texttt{requirements.txt}) to resolve dependencies without polluting the host.
    \item Domain-Specific Build Logic. Manages project-specific compilation and directory structures, incorporating self-healing mechanisms for handling incomplete repositories.
    \item Orchestration Entry Point. Schedules execution modules in topological order to ensure a deterministic transition to a fully operational environment.
\end{enumerate}

The resulting \bashfile serves as the single authoritative artifact for all subsequent environment execution and repair processes. Although this initial configuration may still contain omissions or inconsistencies, it already encodes the project’s inferred dependency structure, build commands, and runtime assumptions in an executable form. All execution attempts, failure diagnoses, and iterative repairs are conducted by modifying and re-running this script. Consequently, \bashfile functions as both the operational backbone of our framework and the final outcome of environment configuration.

\subsection{Test Pyramid Construction}

The multi-agents pipeline iteratively search the \testpyramid, with comprehensive retrieval toolkit equipped. The input of this pipeline is same as \emph{Initial BashFile Generation} (see blue area within Fig.~\ref{fig:overview}), and the output is a list a test commands that in different maturity state. \OurApproach performs a structured five steps to get \testpyramid: \textit{Retrieval}, \textit{Extraction}, \textit{Classification}, \textit{Refinement}, and \textit{Adjustment}.

\textbf{Test Retrieval, Extraction and Classification}. Similarly, in the Retrieval step, agent traverse the repository file tree and search for test-related evidence in both documentation and source code. The search process is guided by the definition of the \testpyramid, which specifies what kinds of tests are relevant at different levels. Using this guidance, we query the knowledge graph to identify relevant files. The output of this step is a set of test-related files, together with their complete textual contents.

For command mining in Extraction step, we process the content of each retrieved file by treating it as one or more text snippets, each associated with the file’s relative path. Each snippet is converted into an extraction prompt that asks the LLM to identify all runnable commands explicitly present in the text (e.g., tests and build commands), without generating new ones. The extracted commands are de-duplicated while preserving order and aggregated into \texttt{involved\_commands}, yielding a unique command list.

 The extracted test commands are filtered and classified according to the definition of \emph{Environment Maturity Hierarchy}. In Classification step, some very unclear and incomplete commands, such as empty, pure comments, and obvious placeholders, are first filtered out. Then the remaining commands are classified. Both the filtering and classification operations from \OurApproach are asked to provide reasons through the thinking process. Finally, the initial \testpyramid are output according to their categories.

\textbf{Test Refinement}. Based on the classification results, the Refinement agent decides whether to return to the Retrieval step for next round of file search. The decision to terminate test collection is made by jointly evaluating the commands collected at each \testpyramid state and the remaining search-round budget. Specifically, the decision criteria are as follow:
\begin{enumerate}
    \sloppy
    \item Test collection accomplished: the search rounds reach the maximum count, OR found $C_{\mathsf{Runnability}}$ and at least one $C_{\mathsf{Testability}}$ and one $C_{\mathsf{Installability}}$.
    \item Test collection not accomplished: No $C_{\mathsf{Runnability}}$, OR found $C_{\mathsf{Runnability}}$ BUT no other $C_{\mathsf{Testability}}$ and no $C_{\mathsf{Installability}}$.
\end{enumerate}
If the test collection task is accomplished, this process stops and moves on to the Adjustment step; otherwise, the workflow returns to the Test Retrieval step for a new round of file search.

\textbf{Test Adjustment}. After classification, we obtain an initial \testpyramid, which may still contain inconsistencies or omissions due to noisy or ambiguous commands. Therefore Adjustment step is used to improve both command quality and coverage. This step addresses common failure modes, including (1) duplicated commands appearing at multiple \testpyramid levels, (2) commands assigned to inappropriate levels due to ambiguous semantics, and (3) missing but commonly required commands for a given level. To resolve these issues, we first remove cross-level duplicates by retaining each command only at its most appropriate level, and then conservatively supplement missing common commands when necessary.

When the level-appropriateness of a command or the presence of typical missing commands is uncertain, the Adjustment agent invokes a \texttt{web search} tool using a constrained query template that includes the level definition, the command under evaluation, and two explicit questions: “Is this command appropriate for this level?” and “What common commands are missing?” Each tool response provides brief reasoning that summarizes both filtering heuristics and web-search evidence. Finally, the agent integrates all such feedback to refine the initial \testpyramid and produce the final one.

\begin{figure}[t]
  \centering
\includegraphics[width=\columnwidth]{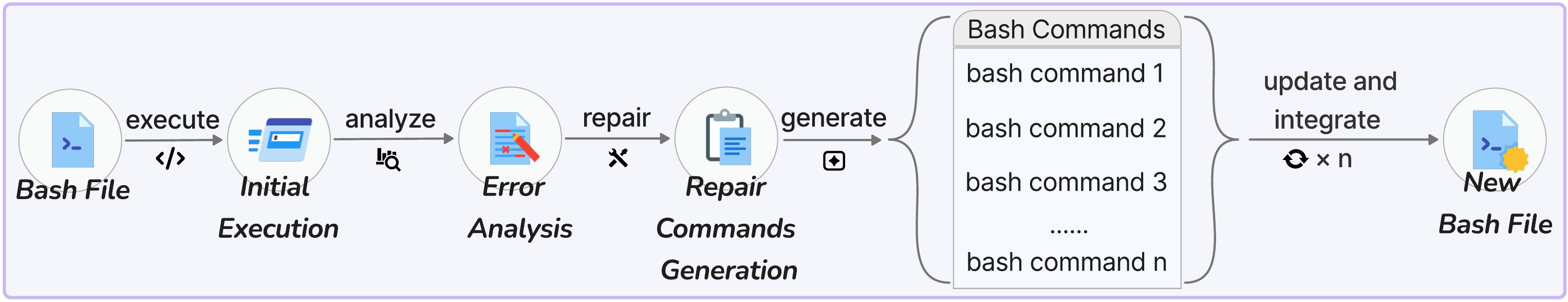}
  \caption{Detailed process of \bashfile Repair. The process iterates through: (1) initial execution to capture runtime errors; (2) analyzing the runtime errors; (3) generating single candidate bash command for repair; and (4) merging and integrating these commands into an updated \bashfile for re-validation.}
  \label{fig:repair}
\end{figure}

\subsection{Interactive Environment Deployment}

The previous steps of \OurApproach outputs two artifacts: a \testpyramid and a global \bashfile. These artifacts serve as the inputs to the \textit{Interactive Environment Deployment} (see green area within Fig.~\ref{fig:overview}), during which the system iteratively executes and repairs the environment configuration. The Interaction Phase consists of two execution cycles: the \textit{Execution Loop} and the \textit{Feedback Loop}. The multi-agent framework repeatedly executes the \bashfile and selected test commands from \testpyramid, analyzes execution feedback, and incrementally updates the \bashfile through a \emph{\bashfile Repair} mechanism. The objective is to derive an executable \bashfile that can independently instantiate a runnable sandbox container. Thus, this interaction step seeks either to elevate the environment to the highest attainable maturity level by incrementally repairing the \bashfile, or to identify the maturity state at which execution stabilizes.

\subsubsection{\textit{Execution Loop}}

The Execution Loop is the entry point of the Interactive Environment Deployment. The objective is to verify whether the current \bashfile can be executed successfully. The agent executes the \bashfile in a containerized environment using the \emph{Docker SDK} and evaluates success based on the returned exit code. If execution fails, the agent analyze the execution result and start \emph{\bashfile Repair}. The Execution Loop iterates until the \bashfile executes successfully or a maximum recursion limit reached. Upon success, the workflow turns to the Feedback Loop. The Execution Loop establishes a stable and repeatable execution context that serves as the foundation for subsequent validation.

\subsubsection{\textit{Feedback Loop}}

If the Execution Loop successfully execute \bashfile, the Feedback Loop is triggered. Its objective is to progressively validate and advance the environment’s maturity by executing test commands in the \testpyramid. The inputs to this loop are the validated \bashfile and a set of candidate test commands annotated with their maturity levels (\maturityI, \maturityII, \maturityIII). The agent first selects a test command based on the current maturity, prioritizing low-risk commands at lower maturity levels and introducing test commands in higher state as the environment stabilizes. The selected test command is then executed within the container and execution output are collected.

Successful execution of a test command may trigger a maturity state transition. If the executed command is from $C_{\mathsf{Installability}}$ or $C_{\mathsf{Testability}}$, the environment is advanced to the $C_{\mathsf{Testability}}$ and $C_{\mathsf{Runnability}}$ , respectively. Only successful execution of $C_{\mathsf{Runnability}}$ indicates that the environment has reached the final target state. As formulation in Section~\ref{sec:problem_statement}, the state-aware transition function $\delta_s$ is instantiated by a prompt-driven agent, which observes the current maturity state, the associated \testpyramid, and the execution feedback of attempted commands. Test selection and state progression are adaptive rather than strictly monotonic. $\delta_s$ decides whether to repeatedly execute commands at the same maturity level, advances to a higher-level, or temporarily revert to lower-level commands if higher-level executions fail persistently, in order to re-establish a stable execution baseline.

When a test command fails, the \emph{\bashfile Repair} workflow is triggered. After repair, the system rolls back to the \bashfile execution step in the Execution Loop to re-validate environment executability. This rollback mechanism mitigates unobserved environment state drift that may be introduced by test commands with side effects (e.g., installation or build operations) by re-establishing a validated execution baseline.

\subsubsection{\textit{BashFile Repair}}

Both the Execution Loop and the Feedback Loop rely on a unified \emph{\bashfile Repair} mechanism (see Fig.~\ref{fig:repair}), which treats the \bashfile as the sole persistent carrier of environment state. All repair actions are expressed as modifications to the \bashfile rather than isolated command executions.

Each repair iteration follows three steps: (1) analyzing execution outputs to identify environment-related failures (e.g., missing dependencies, absent environment variables, compilation or runtime errors); (2) generating candidate repair commands; and (3) integrating selected commands into the Bash File by replacing faulty commands or appending new ones. The updated \bashfile is then re-validated in the Execution Loop. Unlike prior approaches that rely solely on step-by-step command execution~\cite{bouzenia2025you, hu2025repo2run} or wholesale script regeneration~\cite{kovrigin2025piper}, \OurApproach adopts a hybrid strategy. To the best of our knowledge, we are the first to use holistic script maintenance with fine-grained, single-command exploration. By persisting configuration decisions within the \bashfile, this mechanism avoids context loss associated with ephemeral command execution and non-interactive container shells. 

Together, Execution Loop and Feedback Loop constitute a double closed-loop Environment Repair process, while the \bashfile Repair mechanism provides a unified and sustainable repair basis for the two closed loops. By explicitly concentrating the execution status in the \bashfile, the system can steadily advance between different maturity stages and effectively avoid the problem of context loss in single-command exploration in complex environments.

%% file: sections/4.ExperimentalSetup.tex
Our evaluation focuses on the following research questions:
\begin{itemize}
    \item \textbf{RQ1 (Effectiveness)}: How effective is \OurApproach at establishing environments and executing tests compared to state-of-the-art approaches?
    \item \textbf{RQ2 (Test Pyramid)}: How does the proposed \testpyramid organize test usage during automated deployment to provide meaningful signals of environment maturity?
    \item \textbf{RQ3 (Ablation Study)}: How do different repair components of \OurApproach contribute to successful environment deployment and guide the repair process?
\end{itemize}

\subsection{Benchmarks}

Our experiments evaluate \OurApproach on four benchmarks. To our knowledge, this is one of the most comprehensive evaluations of automated environment setup to date, covering all publicly available benchmarks.

Specifically,  \OurApproach is evaluated on the following four benchmarks: \textit{ExecutionAgent-Bench}~\cite{bouzenia2025you}, which includes 50 open-source projects spanning 14 programming languages and verifies results by reproducing CI/CD test logs; \textit{EnvBench-Python}~\cite{eliseeva2025envbench}, consisting of 96 curated Python repositories with Pyright-based setups; \textit{Repo2Run-Bench}~\cite{hu2025repo2run}, which contains 420 instances in total, of which we evaluate the same subset of 122 instances used in prior work (PIPER~\cite{kovrigin2025piper}); and \textit{Installamatic-Bench}~\cite{milliken2025beyond}, comprising 40 curated Python projects with tests located in a \texttt{tests} directory.

\subsection{Baselines}

We consider the following relevant related works and LLMs as our baselines to achieve a comprehensive comparison.

\begin{enumerate}
\item \text{ExecutionAgent}~\cite{bouzenia2025you} leverages meta-prompting to retrieve language-specific guidelines and iteratively refines installation commands based on system feedback.

\item \text{Repo2Run}~\cite{hu2025repo2run} generates Dockerfiles and employs rollback mechanisms to prevent environment contamination. Runtime verification is performed using \textit{pytest}.

\item \text{PIPER}~\cite{kovrigin2025piper} trains a lightweight language model using a two-stage training for automated environment deployment.

\item \text{Installamatic}~\cite{milliken2025beyond} extracts installation instructions from repository artifacts and applies a build-and-repair pipeline to generate functional Dockerfiles.

\item \text{LLM baselines}: closed-source OpenAI models (GPT-5, GPT-4o, and GPT-4o-mini) and open-source Qwen3 models (8B, 14B, and 32B).

\end{enumerate}

\subsection{Methodology for RQ1}
\textbf{Evaluation protocol}. We report results using \textit{Pass@k}, where the value of 
$k$ follows the configuration used in each benchmark’s prior work. For fair comparison, we adopt the same \textit{Pass@k} settings as the corresponding baselines.

Specifically, EnvBench-Python~\cite{eliseeva2025envbench}.
Baselines are evaluated under a zero-shot setting with $k=5$. A repository is considered successful if at least one attempt exits with code 0 and reports no issues under \textit{Pyright}. For fairness, \OurApproach applies the same Pyright-based check and reports results from a single agentic run.
Repo2Run-Bench~\cite{hu2025repo2run}.
Baselines use $k=5$, where success requires completing \textit{pytest} test collection without errors; Repo2Run additionally requires the Dockerfile to build successfully. \OurApproach reports the number of repositories reaching \maturityII. Results for \maturityI and \maturityIII are reported separately.
ExecutionAgent-Bench~\cite{bouzenia2025you}.
A configuration is considered successful if the project can be built and the test suite executed. All results are reported with $k=5$. The higher maturity states of \OurApproach are analyzed separately.
Installamatic-Bench~\cite{milliken2025beyond}.
Success is defined as building the Dockerfile and executing at least one test suite. Baselines follow the original setting with $k=10$, while \OurApproach reports results using $k=3$.
We additionally evaluate sampled C/C++ repositories under the same test-based success criteria.

\input{sections/tables/table-effectness}

\begin{figure}[t]
  \centering
  \includegraphics[width=\columnwidth]{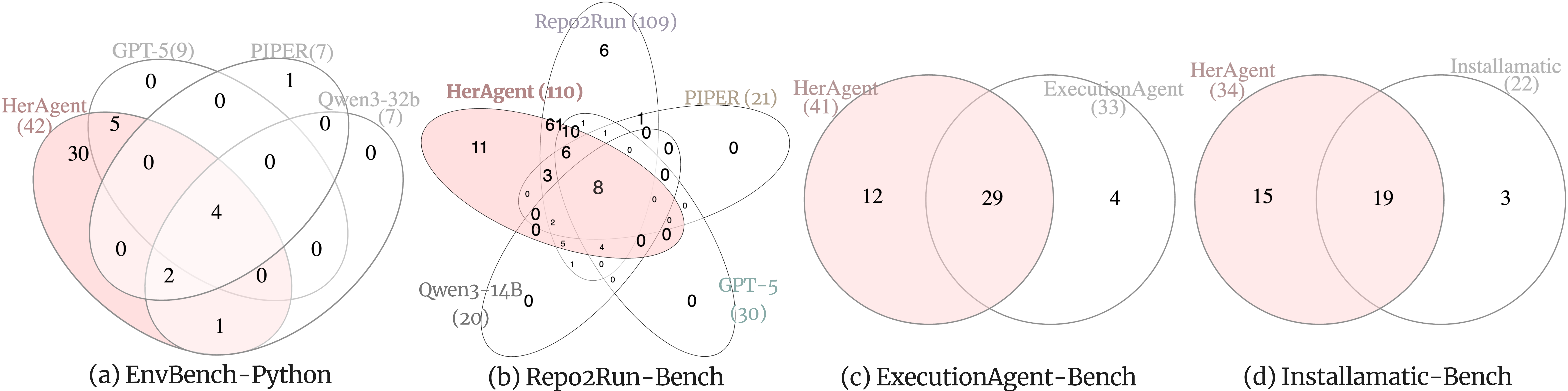}
  \caption{Venn diagram showing the complement and uniqueness of \OurApproach on four benchmarks: EnvBench-Python, Repo2Run-Bench, ExecutionAgent-Bench and Installamatic-Bench.}
  \label{fig:venn}
\end{figure}

\subsection{Methodology for RQ2}
We design an empirical study to investigate how our approach extracts a test pyramid from software repositories and how the corresponding tests are executed during the interactive environment deployment process. The study is conducted on three benchmarks: ExecutionAgent-Bench~\cite{bouzenia2025you}, Repo2Run-Bench~\cite{hu2025repo2run}, and Installamatic-Bench~\cite{milliken2025beyond}. It consists of three complementary analyses: \testpyramid construction, interactive test execution, and environment maturity state transitions. Together, these analyses provide a comprehensive view of how our approach executes tests and how projects progress through different environment readiness states.

First, we analyze all \testpyramid obtained after the Tests Adjustment step across the three benchmarks. By examining their distribution, we characterize the number of test commands at different pyramid levels, which allows us to describe the structural complexity of the test suites. Second, after the test selection step, we analyze the execution outcomes of test commands across different maturity states, thereby assessing the relative execution difficulty of different types of tests. Finally, we conduct a case-study–based analysis to illustrate how test selection decisions influence environment maturity transitions during interactive deployment.

\subsection{Methodology for RQ3}
To evaluate the effectiveness of the \emph{\bashfile Repair} mechanism, we conduct a set of ablation experiments isolating key components of the repair pipeline. "Whole Script Repair" refers to regenerating the entire \bashfile in response to failures, "Single Command Repair" refers to generating and executing individual bash commands in isolation, and "Interactive Feedback" refers to the iterative loop in which the agent observes execution errors and applies corrective actions. Because "Whole Script Repair" and "Single Command Repair" carry different amounts of semantic information per repair turn, we set the recursion limit to 250 for experiments with only "Single Command Repair" (without "Whole Script Repair"), and 200 for all other experiments. All experiments are performed on the same \bashfile and \testpyramid, using the ExecutionAgent-Bench~\cite{bouzenia2025you} and evaluated with the Pass@1 metric.

\subsection{Implementation Details}

\OurApproach is evaluated using a fully agentic workflow that operates over a graph-structured control flow implemented with the LangGraph~\footnote{\url{https://www.langchain.com/langgraph}} framework. All agent decisions are powered by the GPT-5 model (\texttt{gpt-5-2025-08-07})~\footnote{\url{https://platform.openai.com/docs/models/gpt-5}}, with a fixed decoding configuration and temperature set to 1 across all experiments. This setting encourages sufficient exploration during multi-step environment reasoning while ensuring consistency across benchmarks. Each agent step is composed of one or more subgraphs. We enforce explicit recursion limits to bound the maximum number of graph execution iterations for each subgraph component. Specifically, the recursion limits are set to 100 for \emph{Initial BashFile Generation}, 200 for \emph{Test Pyramid Construction}, and 200 for \emph{Interactive Environment Deployment}. These limits prevent infinite repair loops yet allow sufficient exploration to resolve complex dependency and configuration errors, thereby ensuring workflow termination and reproducibility.

%% file: sections/tables/table-effectness.tex
\begin{table}[t]
  \centering
  \footnotesize
  \renewcommand{\arraystretch}{0.9}
  \setlength{\tabcolsep}{2pt}
  \caption{Effectiveness comparison between \OurApproach and related works across four benchmarks.}
  
  \label{tab:comparison-4grid}
  \makebox[0.85\linewidth][l]{%
    \footnotesize
    \textsuperscript{$\bigstar$} Evaluation with \maturityII in \testpyramid ;\;
    \textsuperscript{$\dagger$} Evaluation with Pytest ;\;
  }
  \vspace{0.5em}
  \makebox[0.85\linewidth][l]{%
    \footnotesize
    \textsuperscript{$\ast$} Evaluation with Pyright ;\;
    \textsuperscript{$\ddagger$} Evaluation with test suites ;\;
    ---  Data unavailable.
  }
  \vspace{-0.3em}

  \begin{subtable}[t]{0.46\columnwidth}
    \raggedleft
    \captionsetup{justification=raggedleft,singlelinecheck=false}
    \caption{Comparison on EnvBench-Python}
    \begin{tabular}{l|lcccc}
      \toprule
      \textbf{Metric} & \textbf{Approach} & & & & \textbf{\# Success} (out of 96) \\
      \midrule
      
      \multirow{7}{*}{Pass@5}
      & GPT-5 \textsuperscript{$\ast$}         & & &     & 9 \\
      & GPT-4o \textsuperscript{$\ast$}         &&&     & 7 \\
      & GPT-4o-mini \textsuperscript{$\ast$}    &&&     & 5 \\
      & Qwen3-32B \textsuperscript{$\ast$}     &&&   & 8 \\
      & Qwen3-14B \textsuperscript{$\ast$}     &&&   & 5 \\
      & Qwen3-8B \textsuperscript{$\ast$}      &&&   & 2 \\
      & PIPER \textsuperscript{$\ast$}        &&&   & 8 \\
      \midrule
      \cellcolor{tablelightblue}Pass@1 & \cellcolor{tablelightblue}\OurApproach \textsuperscript{$\ast$} &\cellcolor{tablelightblue}&\cellcolor{tablelightblue}&  \cellcolor{tablelightblue} & \cellcolor{tablelightblue}\textbf{42} \\
      \bottomrule
    \end{tabular}
  \end{subtable}
  \hfill
  \vspace{0.6em}
  \begin{subtable}[t]{0.43\columnwidth}
    \raggedright
    \captionsetup{justification=raggedright,singlelinecheck=false}
    \caption{Comparison on Repo2run Dataset}
    \begin{tabular}{l|lcccc}
      \toprule
      \textbf{Metric} & \textbf{Approach} & & & & \textbf{\# Success} (out of 122) \\
      \midrule
      
      \multirow{7}{*}{Pass@5}
      & GPT-5 \textsuperscript{$\dagger$}          & & &     & 30 \\
      & GPT-4o \textsuperscript{$\dagger$}         &&&     & 16 \\
      & GPT-4o-mini \textsuperscript{$\dagger$}    &&&     & 17 \\
      & Qwen3-32B \textsuperscript{$\dagger$}     &&&   & 19 \\
      & Qwen3-14B \textsuperscript{$\dagger$}     &&&   & 20 \\
      & Qwen3-8B \textsuperscript{$\dagger$}      &&&   & 7 \\
      & PIPER \textsuperscript{$\dagger$}        &&&   & 28 \\
      \midrule
      
        --- & Repo2Run \textsuperscript{$\dagger$}        &&&   & 109 \\
      \midrule
      \cellcolor{tablelightblue}Pass@3 & \cellcolor{tablelightblue}\OurApproach \textsuperscript{$\bigstar$}    &\cellcolor{tablelightblue}&\cellcolor{tablelightblue}& \cellcolor{tablelightblue}  & \cellcolor{tablelightblue}\textbf{110} \\
      \bottomrule
    \end{tabular}
  \end{subtable}
  \hfill
  \vspace{0.6em}
  \begin{subtable}[t]{0.46\columnwidth}
    \raggedleft
    \captionsetup{justification=raggedleft,singlelinecheck=false}
    \caption{Comparison on ExecutionAgent Datast}
    \begin{tabular}{l|lcccc}
      \toprule
      \textbf{Metric} & \textbf{Approach} & & & & \textbf{\# Success} (out of 50) \\
      \midrule

      \multirow{3}{*}{Pass@3}
      & LLM script \textsuperscript{$\ddagger$}      &&&   & 5 \\
      & AutoGPT \textsuperscript{$\ddagger$}         &&&   & 4 \\
      & ExecutionAgent \textsuperscript{$\ddagger$}   &&&   & 33 \\
      \midrule
      \cellcolor{tablelightblue}Pass@3 & \cellcolor{tablelightblue}\OurApproach \textsuperscript{$\bigstar$}     &\cellcolor{tablelightblue}&\cellcolor{tablelightblue}&  \cellcolor{tablelightblue} & \cellcolor{tablelightblue}\textbf{41} \\
      \bottomrule
    \end{tabular}
  \end{subtable}
  \hfill
  \vspace{0.6em}
  \begin{subtable}[t]{0.49\columnwidth}
    \raggedright
    \captionsetup{justification=raggedright,singlelinecheck=false}
    \caption{Comparison on Installamatic Datast}
    \begin{tabular}{l|lcccc}
      \toprule
      \textbf{Metric} & \textbf{Approach} & & & & \textbf{\# Success} (out of 40) \\
      \midrule

      Pass@10 & Installamatic \textsuperscript{$\dagger$}      &&&   & 21 \\
      \midrule
      \cellcolor{tablelightblue}Pass@3 & \cellcolor{tablelightblue}\OurApproach \textsuperscript{$\bigstar$}     &\cellcolor{tablelightblue}&\cellcolor{tablelightblue}&  \cellcolor{tablelightblue} & \cellcolor{tablelightblue}\textbf{34} \\
      \bottomrule
    \end{tabular}
  \end{subtable}
  \vspace{1em}

\end{table}

%% file: sections/5.ExperimentResults.tex
\subsection{Results for RQ1 (Effectiveness)}

Table~\ref{tab:comparison-4grid} summarizes the results across four representative benchmarks. Across all benchmarks, \OurApproach achieves state-of-the-art performance, consistently configuring the largest number of repositories. Notably, these gains are achieved with fewer attempts than most baselines, indicating that the improvements come from stronger reasoning and execution robustness rather than larger sampling budgets.

\textbf{State-of-the-art performance across all considered benchmarks under stricter evaluation criteria}.
\OurApproach achieves state-of-the-art performance across all considered benchmarks.
Specifically, on EnvBench-Python, \OurApproach successfully configures 42/96 repositories in a single attempt, outperforming the strongest baseline (9/96) evaluated with five attempts, yielding a 4.7× improvement. On Repo2Run-Bench, \OurApproach achieves 110/122 successes under a stronger test-execution criterion, slightly exceeding Repo2Run (109/122) and improving over the best zero-shot baseline (30/122) by +80 successes. On ExecutionAgent-Bench, \OurApproach resolves 41/50 projects, surpassing ExecutionAgent (33/50) by +8 projects. On Installamatic-Bench, \OurApproach achieves 34/40 successes, compared to 21/40 by Installamatic, corresponding to a +62\% relative improvement.

\textbf{Complementarity and unique coverage.} Fig.~\ref{fig:venn} illustrates the overlap and uniqueness of \OurApproach relative to prior methods across all benchmarks. In each setting, \OurApproach not only shares a substantial number of successful configurations with strong baselines, but also resolves a notable set of instances that none of the baselines can handle. On EnvBench-Python, \OurApproach uniquely solves \texttt{30} repositories that remain unresolved by all other models. On Repo2Run-Bench, \OurApproach achieves \texttt{11 unique} successes beyond those covered by Repo2Run and other baselines, despite their overlapping strengths. Similar patterns are observed on ExecutionAgent-Bench and Installamatic-Bench, where \OurApproach resolves \texttt{12} and \texttt{15} additional configurations, respectively. These results indicate that \OurApproach captures complementary capabilities rather than simply reproducing existing methods.

\textbf{Effectiveness on the most challenging C/C++ project setups.}
C and C++ projects are widely regarded as the most challenging setting for automated environment setup, due to heterogeneous build systems, complex dependency chains, and strict compiler and toolchain constraints. Notably, ExecutionAgent is the only prior work that evaluates environment setup on C/C++ projects~\cite{bouzenia2025you}, leaving this challenging domain largely unexplored.
Table~\ref{tab:ablation-C++} compares \OurApproach with ExecutionAgent on 24 representative C/C++ repositories. \OurApproach substantially outperforms ExecutionAgent at both \maturityI (23 vs.\ 16) and \maturityII (20 vs.\ 12). More importantly, \OurApproach is the only approach that consistently reaches \maturityIII, successfully configuring 17 projects to a fully runnable state. These results demonstrate that \OurApproach can handle complex, real-world C/C++ environment setups beyond the capabilities of prior work.

\begin{figure}[t]
  \centering
  \includegraphics[width=\columnwidth]{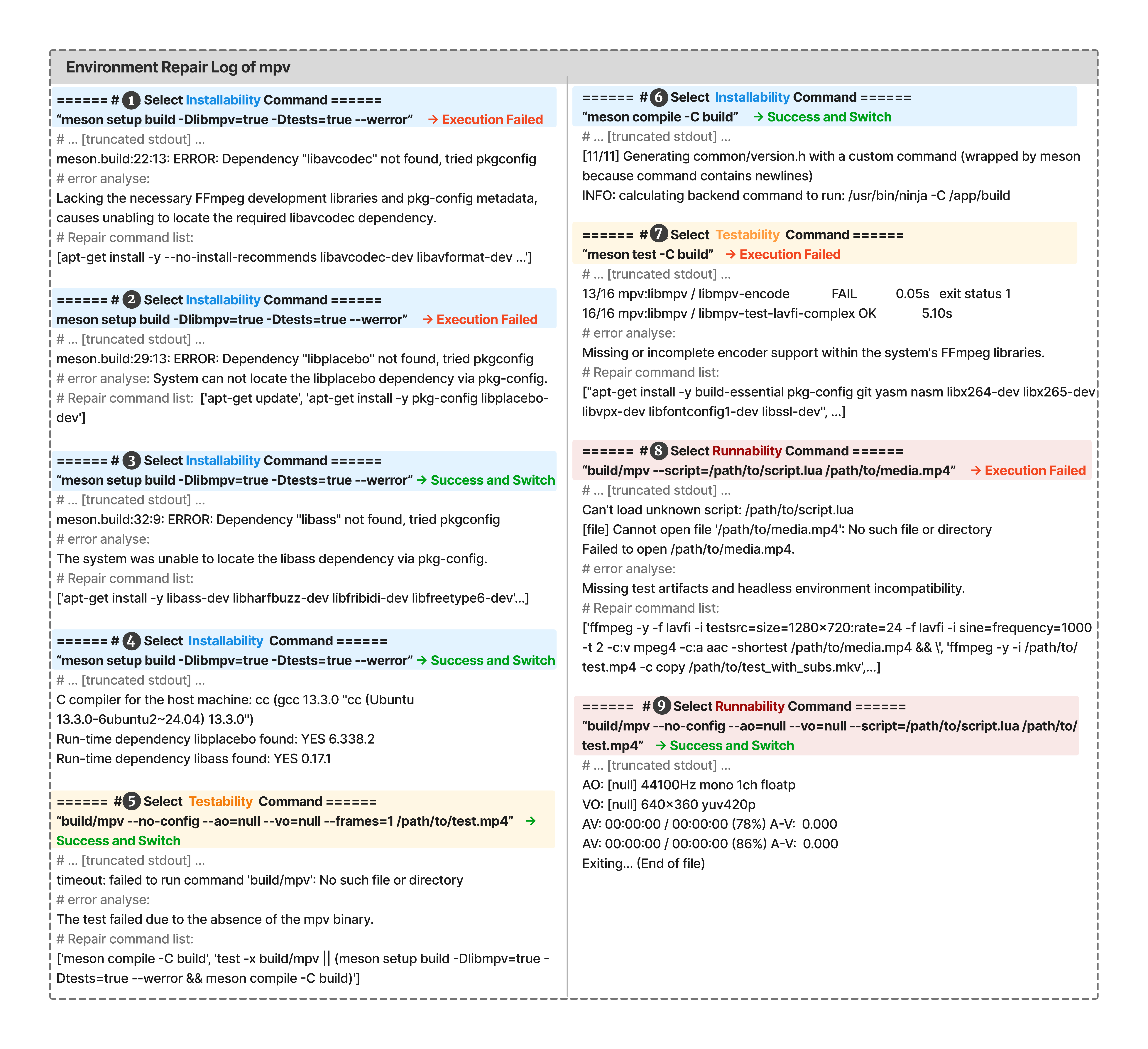}
  \caption{Detailed execution and repair trajectory of \OurApproach on the \emph{mpv} project. The agent selects and executes command in each iteration. Execution failures trigger error analysis and \bashfile repair mechanism, while successes advance the test selection workflow until the environment reaches full readiness.}
  \vspace{-1em}
  \label{fig:detailed_trajectory}
\end{figure}

\textbf{Case study: successful setup of a complex C/C++ project.}
\autoref{fig:detailed_trajectory} presents a representative case in which only \OurApproach successfully configures a complex C/C++ project, \texttt{mpv}, a media player built with the Meson toolchain. The deployment trace exposes a multi-stage dependency chain, where configuration fails sequentially due to missing \texttt{libavcodec}, \texttt{libplacebo}, and \texttt{libass}. \OurApproach resolves these failures by persistently accumulating repair commands in a global script, ensuring that environment state evolves monotonically rather than being overwritten across iterations. This is explicitly confirmed in \textit{\#4}, where previously installed dependencies are recognized by the build system, demonstrating stable preservation of environment context.

Beyond dependency resolution, the trajectory shows that \OurApproach respects the C/C++ build life-cycle, repeatedly re-running configuration after each repair and invoking compilation only when the binary is confirmed missing (e.g., \textit{\#5}). Finally, to reach \maturityIII, \OurApproach identifies and repairs runtime failures caused by missing artifacts by synthesizing required scripts and test media, enabling successful end-to-end execution (\textit{\#9}). This case highlights the key novelty of \OurApproach: holistic, state-preserving environment reasoning that goes beyond error-driven, single-command repairs and enables robust setup of complex real-world C/C++ projects.

\begin{summary}
\textbf{Answer to RQ1:}
Across all benchmarks, \OurApproach  achieves state-of-the-art effectiveness under stricter, execution-aware evaluation criteria while using fewer attempts than prior work.
 \OurApproach also demonstrates strong complementarity, resolving 30, 11, 12, and 15 unique configurations across the four benchmarks. This advantage is most evident on complex C/C++ projects, where  \OurApproach is the only approach that consistently reaches full runnability, successfully configuring 17 projects end to end.
\end{summary}

\begin{figure}[t]
  \centering
  
  \begin{subfigure}{0.32\columnwidth}
    \centering
    \includegraphics[width=\linewidth]{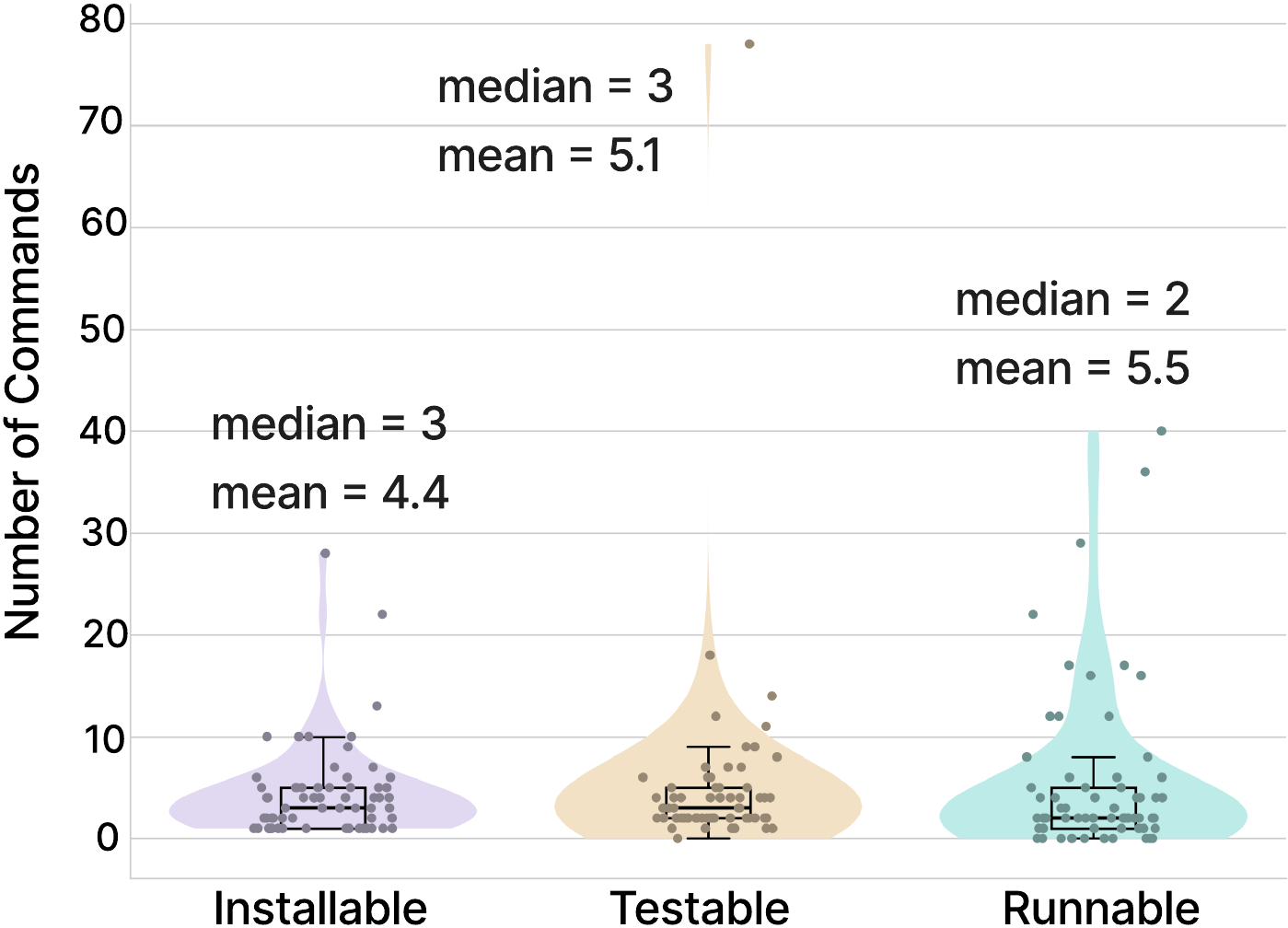}
    \caption{Repo2Run-Bench}
  \end{subfigure}
  \hfill
  \begin{subfigure}{0.32\columnwidth}
    \centering
    \includegraphics[width=\linewidth]{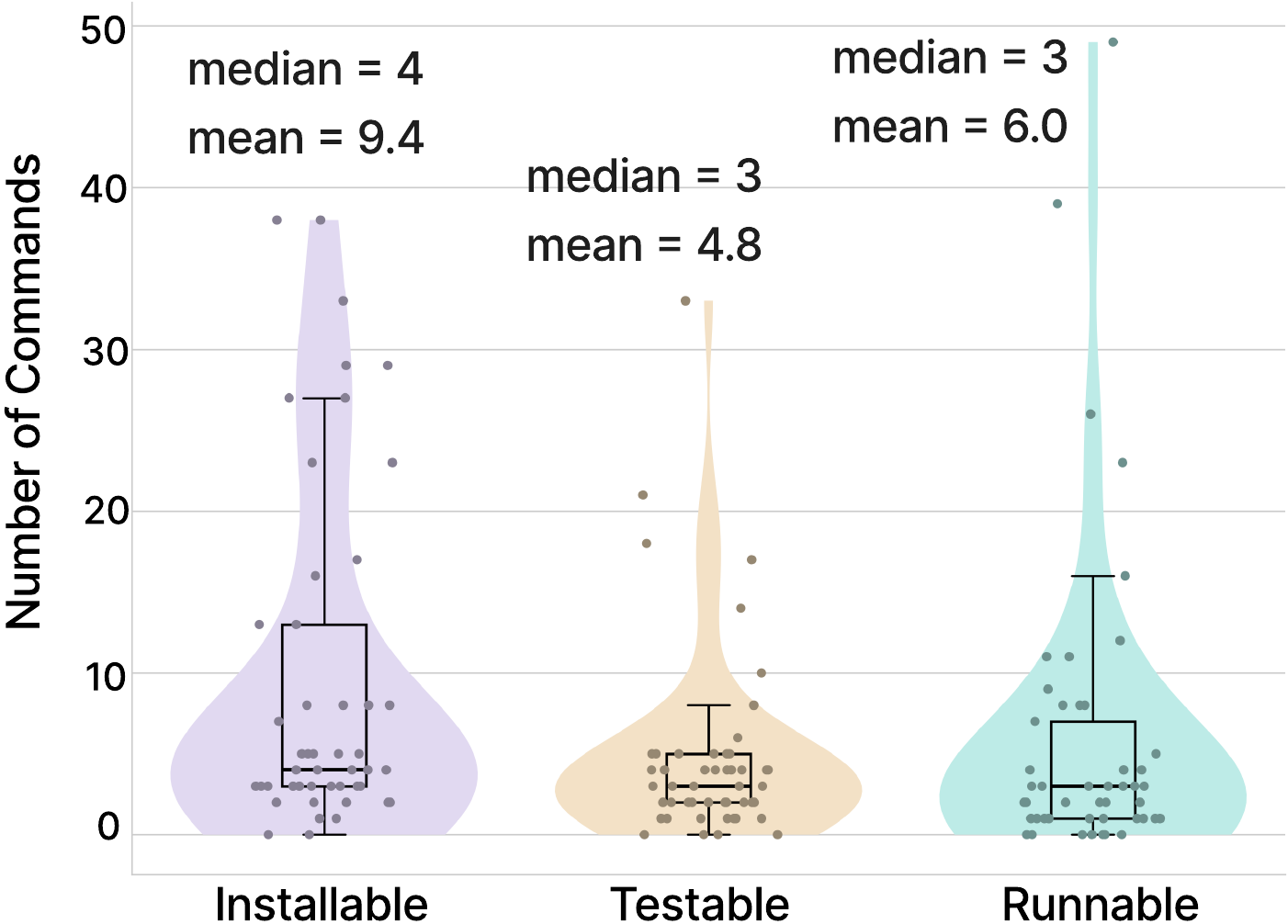}
    \caption{ExecutionAgent-Bench}
  \end{subfigure}
  \hfill
  \begin{subfigure}{0.32\columnwidth}
    \centering
    \includegraphics[width=\linewidth]{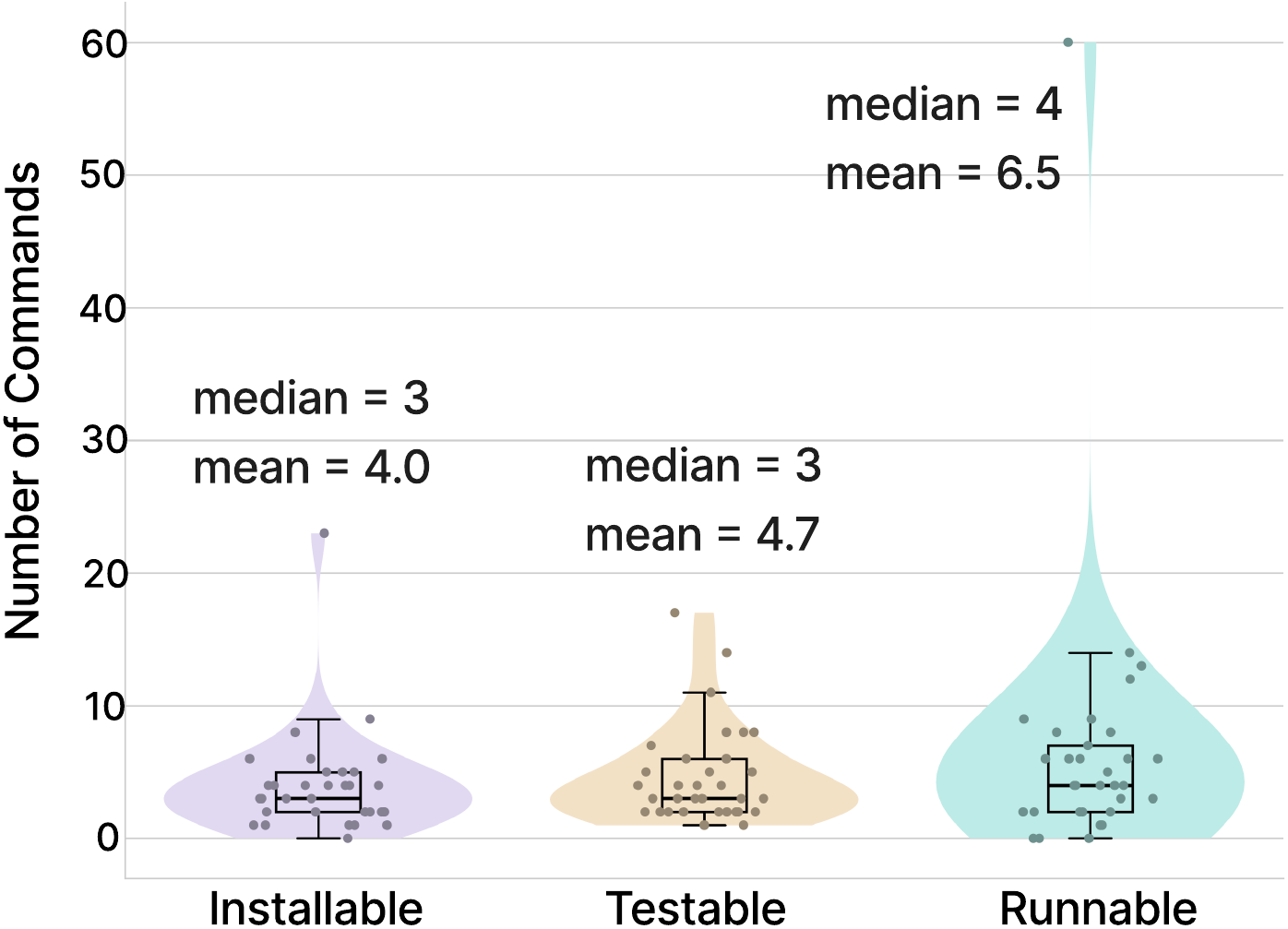}
    \caption{Installamatic-Bench}
  \end{subfigure}

  \vspace{1em}

  \begin{subfigure}{0.32\columnwidth}
    \centering
    \includegraphics[width=\linewidth]{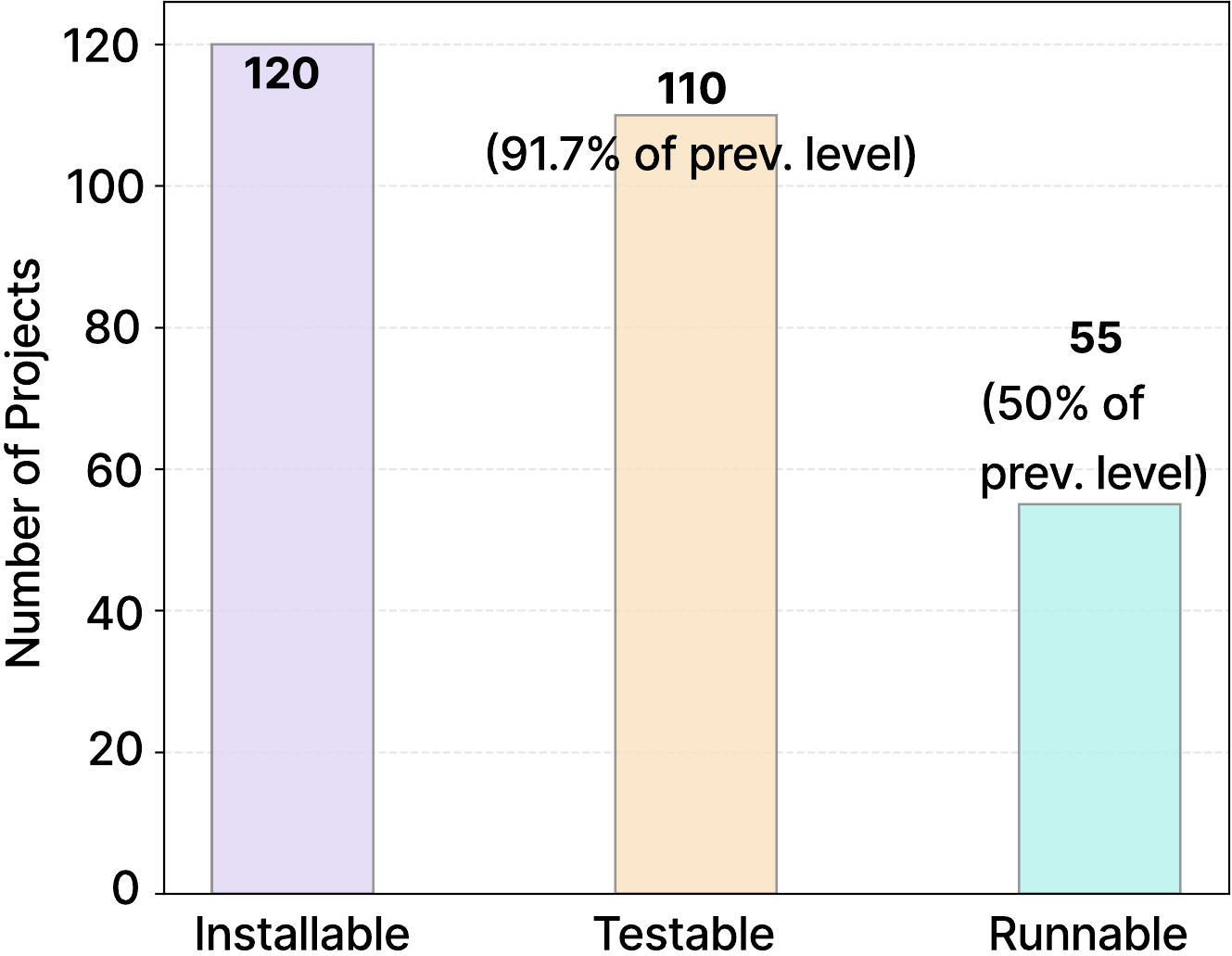}
    \caption{Repo2Run-Bench}
  \end{subfigure}
  \hfill
  \begin{subfigure}{0.32\columnwidth}
    \centering
    \includegraphics[width=\linewidth]{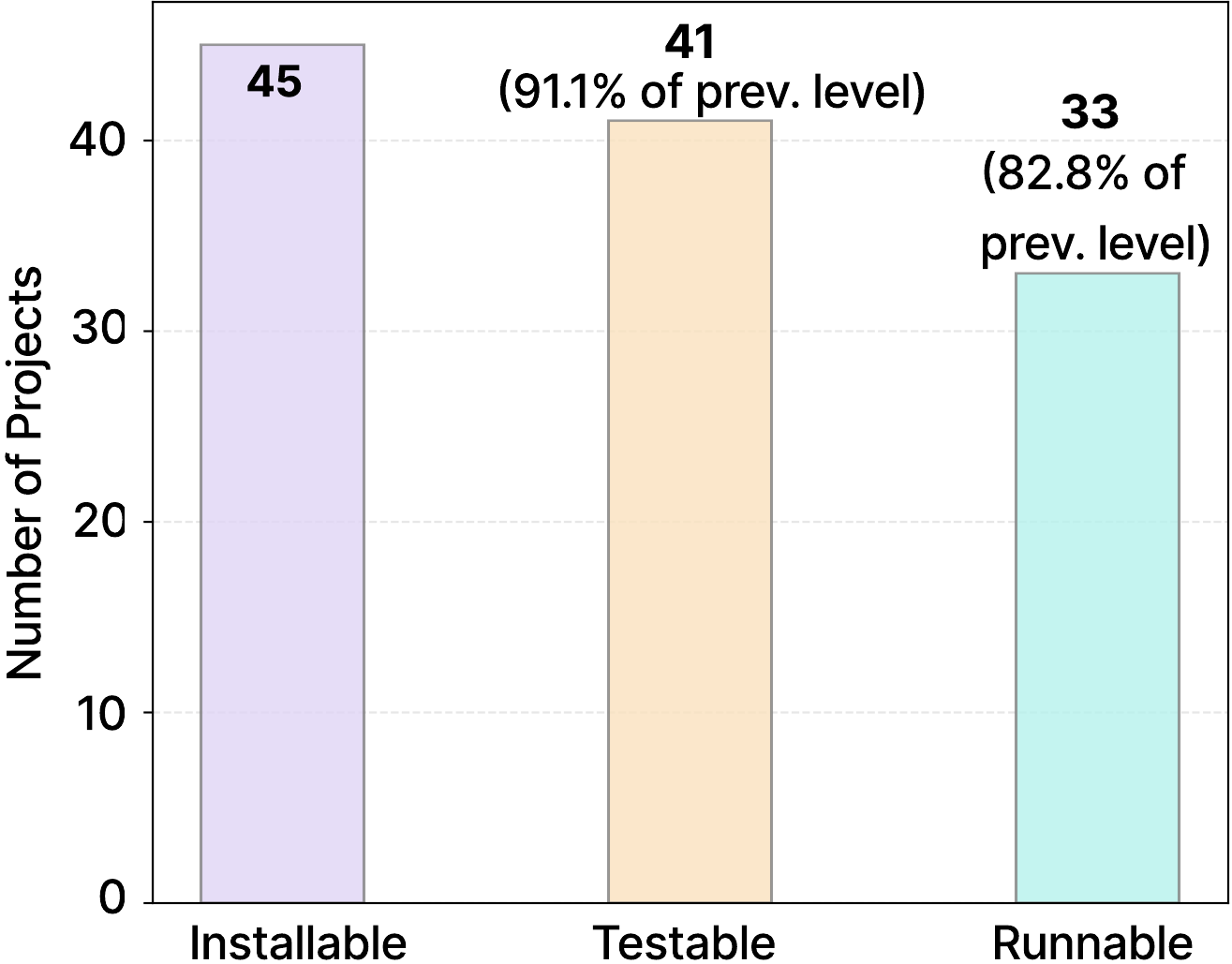}
    \caption{ExecutionAgent-Bench}
  \end{subfigure}
  \hfill
  \begin{subfigure}{0.32\columnwidth}
    \centering
    \includegraphics[width=\linewidth]{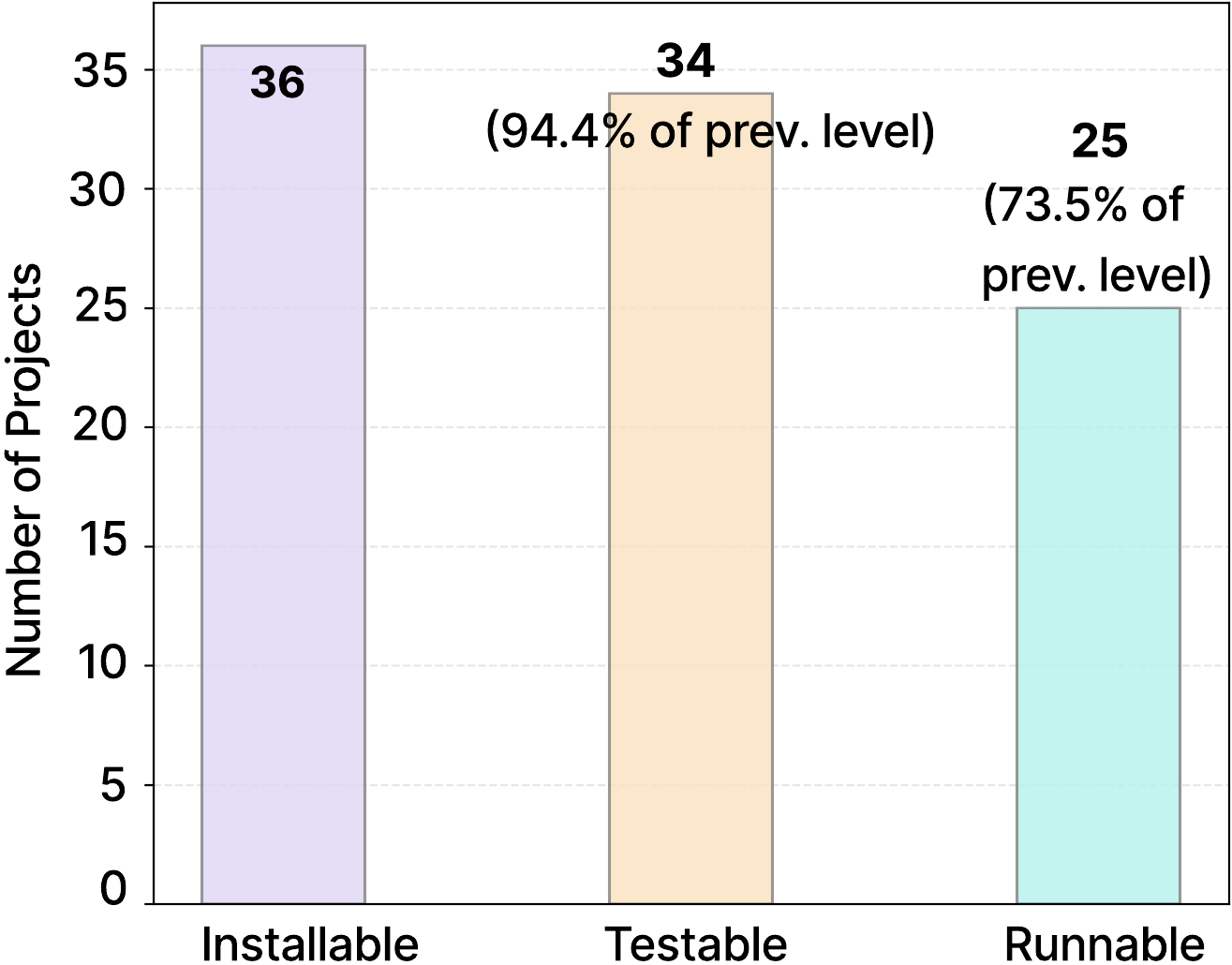}
    \caption{Installamatic-Bench}
  \end{subfigure}
  \vspace{1em}

  \caption{Test Pyramid statistics across benchmarks. (a)(b)(c) Distribution of commands collected for different maturity states. (d)(e)(f) Projects retention rates across different maturity states.}
  \label{fig:test-statistic}
\end{figure}

\subsection{Results for RQ2 (Analysis of Test Pyramid)}
\subsubsection{Statistic of Test Pyramid Collection.}

To analyze the \testpyramid across benchmarks, we visualized command distributions using violin plots overlaid with project-level scatter plots(top panel of Fig.~\ref{fig:test-statistic}). The results reveal a skewed distribution that while the median command counts remain low (typically 2 to 4), the means are consistently higher. This indicates a long-tail effect, meaning that while most projects require simple configuration, a specific subset demands significantly more complex command sequences. Notably, the multilingual ExecutionAgent-Bench(Fig.~\ref{fig:test-statistic}(b)) shows significant outliers in the 'Installable' category, whereas the Python-only Repo2run-Bench(Fig.~\ref{fig:test-statistic}(a)) and Installamatic-Bench(Fig.~\ref{fig:test-statistic}(c)) are more convergent. This suggests that Python’s unified build tools (e.g., pip) provide clear, predictable configuration paths, whereas mixed-language projects lack such standards, resulting in a more chaotic environment. The mean command count for $C_{\mathsf{Runnability}}$ is generally higher than that of $C_{\mathsf{Testability}}$, reflecting the increased complexity of configuring a full execution environment compared to a test suite.

\input{sections/tables/table-ablation-C++}

\begin{figure}[t!]
  \centering
  \includegraphics[width=\columnwidth]{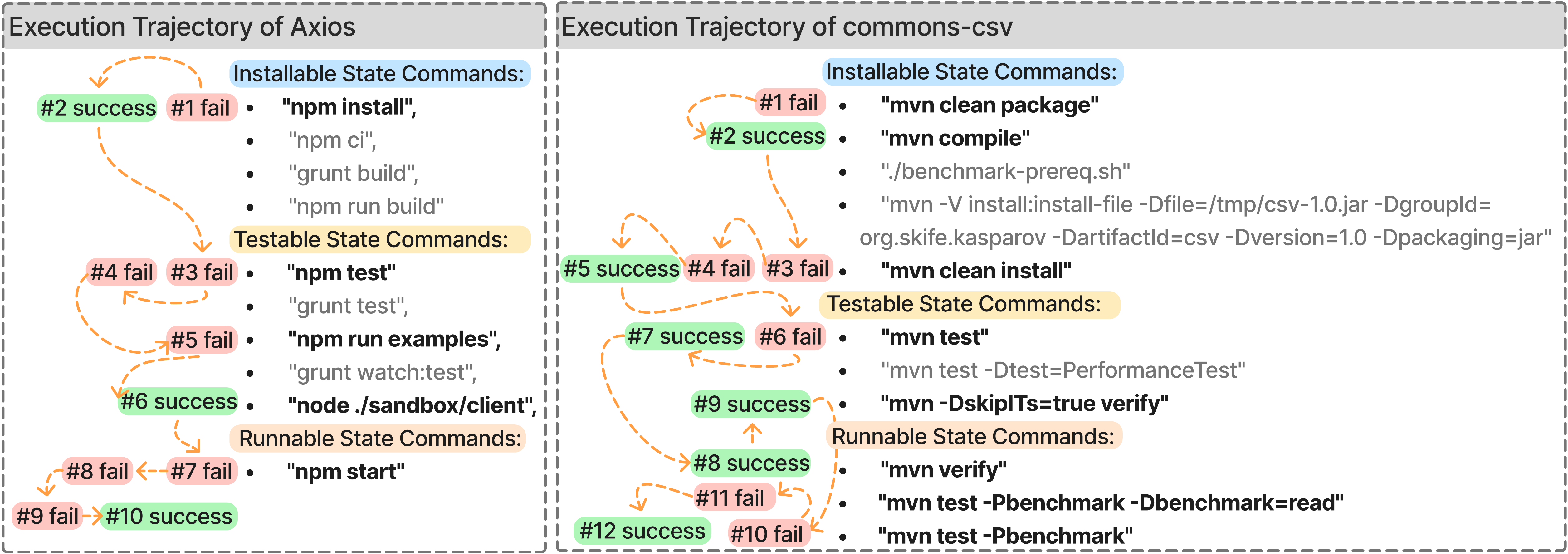}
  \caption{Test Pyramid selection and execution trajectory produced by \OurApproach.}
  \label{fig:test_trajectory}
\end{figure}

\subsubsection{Statistic of Test Pyramid Execution.}

We quantified the project retention rates across different maturity state. As illustrated in the bottom panel of Fig.~\ref{fig:test-statistic}, all benchmarks exhibit a funnel effect, yet the attrition profiles vary significantly between maturity states. 

In contrast, the transition from \maturityII to \maturityIII reveals a sharp divergence in difficulty. \OurApproach on Repo2run-Bench(Fig.~\ref{fig:test-statistic}d) suffered the most substantial drop-off, with the project count halving from 110 to 55 (50\% retention). This suggests that a significant portion of the dataset, while testable, lacks the necessary configuration for full execution. Conversely, on Installamati-Bench(Fig.~\ref{fig:test-statistic}f) and ExecutionAgent-Bench(Fig.~\ref{fig:test-statistic}e) we maintained more robust retention rates of 73.5\% (25 projects) and approximately 80\% (33 projects), respectively. These findings reinforce the distinction in complexity between testing and execution, that while test suites can often run in partially configured environments, full reproducibility requires a comprehensive resolution of system-level dependencies and runtime configurations.

The transition from \maturityI to \maturityII is relatively smooth. All datasets maintained high retention rates: Installamatic-Bench (Fig.~\ref{fig:test-statistic}f) led with 94.4\% (34/36), followed by Repo2run-Bench (Fig.~\ref{fig:test-statistic}d) (91.7\%) and ExecutionAgent-Bench (Fig.~\ref{fig:test-statistic}e) (91.1\%). This suggests that once dependencies are installed, running basic tests is a low-friction step. Our analysis of execution logs supports this. For instance, in Fig.~\ref{fig:test_trajectory} both \textit{Axios} and \textit{commons-csv} failed their first installation attempt (\#1 fail) but succeeded immediately on the second try. This indicates that initial installation hurdles are often minor and easily resolved, allowing projects to quickly enter the testing phase.

In contrast, advancing from \maturityII to \maturityIII is much harder in Fig.~\ref{fig:test-statistic}. Repo2run-Bench saw a sharp drop, retaining only 50\% of its projects (falling from 110 to 55). Installamatic-Bench and ExecutionAgent-Bench maintained more robust retention rates of 73.5\% (25 projects) and approximately 80\% (33 projects), respectively. These findings reinforce the distinction in complexity between testing and execution, that while test suites can often run in partially configured environments, full reproducibility requires a comprehensive resolution of system-level dependencies and runtime configurations. Detailed logs shown in Fig.~\ref{fig:test_trajectory}reveal why this step is difficult
\begin{itemize}
    \item \textbf{Axios:} The trajectory highlights the gap between executing isolated components and the full application. While the sandbox client (\texttt{"node ./sandbox/client"}, \#6) executed successfully early on, the primary entry point \texttt{"npm start"} (\#7-\#10) proved extremely volatile, failing five consecutive times before finally reaching a stable state.
    
    \item \textbf{commons-csv:} Moving beyond basic unit tests revealed significant instability. Although standard tests (\texttt{"mvn test"}, \#6 and \#7) passed relatively quickly, the project struggled with full installation cycles (\texttt{"mvn clean install"} in \#3-\#5 failed twice) and specific runtime profiles. Notably, the benchmark execution (\texttt{"-Pbenchmark"}, \#10, \#11 and \#12) encountered three failures, succeeding only after multiple attempts.
\end{itemize}
These examples confirm that while test suites are often self-contained, full execution demands comprehensive resolution of system-level dependencies and runtime configurations.

\begin{summary}
\textbf{Answer to RQ2:}
Our analysis reveals a sharp retention drop between testing and execution phases, confirming that full runtime reproducibility demands significantly higher configuration completeness than mere test discovery. This validates the necessity of our hierarchical strategy to progressively resolve deep system-level incompatibilities that standard metrics overlook.
\end{summary}

\subsection{Results for RQ3 (Ablation Studies on \bashfile Repair)}

\input{sections/tables/table-ablation_study}

Table~\ref{tab:ablation-script} reports the evaluation results of individual component's contribution within the \bashfile Repair mechanism. The full \bashfile Repair mechanism achieves superior performance across all maturity levels, successfully bringing 26 projects to a \maturityIII and 32 to a \maturityII. The results identify "Interactive Feedback" as the fundamental driver, as its removal results in catastrophic failure (0 successes), confirming that static generation is insufficient for complex environment setup. "Single Command Repair" proves critical for satisfying precise compiler and dependency requirements, as its removal causes a sharper decline in \maturityII (32 to 27) compared to removing "Whole Script Repair" (32 to 31). However, "Whole Script Repair" remains indispensable for resolving global dependency conflicts to reach the \maturityIII; relying solely on single commands drops performance to 19 successes. Ultimately, the peak performance is achieved only through the synergy of these components, where the surgical precision of "Single Command Repair" complements the structural integrity provided by Whole Script Repair.

\begin{summary}
\textbf{Answer to RQ3:}
Ablation studies confirm that the \emph{\bashfile Repair} mechanism relies on the critical synergy between interactive feedback and a hybrid repair strategy to prevent catastrophic failure. Specifically, combining surgical single-command fixes with holistic script maintenance is indispensable for balancing precise dependency resolution with global consistency.
\end{summary}

%% file: sections/tables/table-ablation-C++.tex
\begin{table}[t]
  \centering
  
  \footnotesize
  \renewcommand{\arraystretch}{1}
  \setlength{\tabcolsep}{4pt}
  \caption{Analyze of difficult projects (overall 24 C/C++ projects) between \OurApproach and ExecutionAgent.}
  \label{tab:ablation-C++}

  \begin{tabular}{c|ccc}
    \toprule
    Approach &  Installablity Success & Testablity Success & Runnablity Success\\
    \midrule
    ExecutionAgent & 16 & 12 & -- \\
    \cellcolor{tablelightblue}\OurApproach & \cellcolor{tablelightblue}\textbf{23} &  \cellcolor{tablelightblue}\textbf{20 } & \cellcolor{tablelightblue}\textbf{17} \\
    \bottomrule
  \end{tabular}
\end{table}

%% file: sections/tables/table-ablation_study.tex
\begin{table}[t]
  \centering
  \footnotesize
\renewcommand{\arraystretch}{1.0}
  \setlength{\tabcolsep}{6pt}
  \caption{Ablation Study of \bashfile Repair}
  \label{tab:ablation-script}
  \begin{tabular}{ccc|ccc}
    \toprule
    \multirow{2}{*}{\makecell[t]{Whole\\Script Repair}} &
    \multirow{2}{*}{\makecell[t]{Single\\Command Repair}} &
    \multirow{2}{*}{\makecell[t]{Interactive\\Feedback}} &
    \multicolumn{3}{c}{\textbf{ Pass@1 (out of 50)}} \\

    \cmidrule(lr){4-6}
     &  &  &
    Installability & Testability & Runnability \\
    \midrule

    \checkmark & \checkmark  & \checkmark & \cellcolor{tablelightblue} \textbf{43}& \cellcolor{tablelightblue} \textbf{32}& \cellcolor{tablelightblue} \textbf{26} \\
    \checkmark &  & \checkmark &  39&  27&  21 \\
               & \checkmark & \checkmark &  37&  31&  19 \\
    \checkmark &  &  &  13&  3&  0\\

    \bottomrule
  \end{tabular}
\end{table}

%% file: sections/6.RelatedWorks.tex
\textbf{Automated Environment Setup.} Early automation relies on template-based generators~\cite{microsoft_generator_docker, cloud66_starter} or rule-based Dockerfile synthesizers~\cite{ye2021dockergen, horton2019dockerizeme}. Bridging towards learning-based methods, Rosa et al.~\cite{rosa2023automaticallygeneratingdockerfilesdeep} pioneered the use of deep learning for Dockerfile generation, though primarily focusing on static correctness. In contrast, modern agentic approaches adopt an execute--observe--repair loop \cite{iter}. While general autonomous frameworks like AutoDev~\cite{tufano2024autodevautomatedaidrivendevelopment} and others~\cite{hong2023metagpt,yang2024swe,wang2024opendevin} aim for end-to-end development, they often lack specialized knowledge for complex system dependencies. Domain-specific tools address these gaps through optimization and rigorous validation. PIPER~\cite{kovrigin2025piper} reduces inference costs, while Treefix~\cite{souza2025treefixenablingexecutiontree} enables execution via prefix trees. To ensure reproducibility, RepoST~\cite{xie2025repost} and Repo2Run~\cite{hu2025repo2run} leverage sandbox testing and context mining to construct environments at the repository level. For compiled languages, CXXCrafter~\cite{10.1145/3729386} specifically tackles the intricacies of C/C++ build systems and dependency management, surpassing general agents in build success rates.

To evaluate these capabilities, benchmarks have evolved from simple bug reproduction~\cite{guo2025swe} to specialized environment assessments. CSR-Bench~\cite{xiao-etal-2025-csr} targets the deployment of scientific research software, while DI-BENCH~\cite{zhang2025dibenchbenchmarkinglargelanguage} isolates the challenge of dependency inference. Addressing the need for scalable and dynamic testbeds, R2E~\cite{pmlr-v235-jain24c} converts arbitrary GitHub repositories into executable environments, further extended by R2E-Gym~\cite{jain2025r2egymproceduralenvironmentshybrid} for procedural generation. Despite massive scaling efforts like Deploy-Master~\cite{wang2026deploymasterautomatingdeployment50000}, which automates the deployment of over 50,000 tools, the field still lacks a unified, execution-driven standard for verifying environment readiness across diverse languages.

\textbf{Test-suites as Oracles.} Regression testing research increasingly treats test executions as \textit{probes} for system state rather than simple checks \cite{regression}. To assess correctness without explicit assertions, approaches like TOGA~\cite{dinella2022toga} and TOGLL~\cite{hossain2024togll} generate oracles using neural models and LLMs.
Beyond oracle generation, the structural organization of validation is fundamental. The \textit{Test Pyramid} model~\cite{vocke2018practical} establishes a taxonomy for this organization, advocating for a layered distribution—from high-volume, low-latency unit tests to sparse, high-fidelity end-to-end scenarios—to balance feedback speed and diagnostic depth.
However, structural layering alone is insufficient if the interactions between layers are neglected. Derakhshanfar et al.~\cite{Derakhshanfar_2023} highlight that high code coverage in isolation often fails to capture integration faults. Optimization strategies also play a vital role in observation efficiency: DeepOrder~\cite{sharif2021deeporder} prioritizes tests based on historical failure prediction, and Pan et al.~\cite{pan2022test} provide a comprehensive review of such selection methods. notably, Wang et al.~\cite{wang2024hierarchy} demonstrate that \textit{hierarchy-aware} prioritization—leveraging dependency structures—significantly improves efficiency. 

Inspired by this hierarchy-aware principle, we organize environment validation commands along a dependency hierarchy (from static installation to dynamic runtime). This allows our agent to infer environment readiness under uncertainty, transferring the logic of test prioritization to environment configuration.